\documentclass[aps,showpacs,twocolumn,amsmath,amssymb]{revtex4}
\usepackage{graphicx}% Include figure files
\usepackage{dcolumn}% Align table columns on decimal point
\usepackage{bm}% bold math
\begin{document}

\title{ The new explanation of cluster synchronization in the generalized  Kuramoto system}
\author{Guihua Tian$^{1,2}$,\ \ Songhua Hu$^{1,2,3}$,\ \ Shuquan Zhong$^{2}$,\ \ \email{tgh-2000@263.net, husonghua@126.com }}
 \affiliation{$^1$School of Science, Beijing
University of Posts And Telecommunications. Beijing 100876, China.}
 \affiliation{$^2$State Key Laboratory of Information Photonics and Optical
Communications, \\ Beijing University of Posts And Telecommunications.
 Beijing 100876, China.}
 \affiliation{$^3$ School of Electronic and Information Engineering, \\ North China Institute of Science and Technology,Yanjiao 065201, China}
\begin{abstract}

The cluster synchronization (CS) is a very important characteristic for the higher harmonic coupling  Kuramoto system. A novel transformation  is provided, and it gives CS by  the periodic properties of the density function. The periodic properties of the density function also make the cluster sections' boundaries barrier-like, which helps to explain the sensitiveness of CS on the initial conditions of the oscillators. Detailed numerical studies confirm the theoretical predictions from this new view of the symmetry transformation. The work is very beneficial to the  further study on CS in various systems.
\end{abstract}
\maketitle

\section{Introduction}

The symmetries play important role in varies branches of theoretical physics, both in classical and modern areas.
For example, in the classical mechanics, the Kepler problem is easily solved if one unified the conservation of energy, angular momentum which are the results of the symmetry of the Kepler problem. Even without solving the problem, the conservation of the angular momentum will tell one many information, like the motion of the planet in the sun system being of plane, etc. In the paper, we will exploit the symmetry method to study the generalized Kuramoto system and give the answer to the question of cluster synchrony state without solving the problem directly.

The Kuramoto model (KM) captures the main property of the collective synchronization with the first harmonic coupling as $H(\theta_j-\theta_i)=K_{ij}\sin (\theta_j-\theta_i) $ and revealed the second continuous transition at the critical coupling strength $K_c$. KM is applied in many physical, biological and social systems, including electrochemical oscillators, Josephson junction arrays,  cardiac pacemaker cells, circadian rhythms in mammals, network structure and neural network\cite{kura}-\cite{gupta}.

KM have been generalized in many aspects\cite{gupta}-\cite{ping}.  one of them is the introduction of the globally higher  harmonic coupling $H(\theta_j-\theta_i)=K_{ij}\sin m(\theta_j-\theta_i),m\in N, \ m>1 $,  where many new and interesting phenomena appear, like the cluster synchronization (CS) , and switching of the oscillators between different clusters with
the external force, etc.\cite{ott}-\cite{niyo}. Higher  harmonic coupling (HHC) is dominating in $\phi$-Josephson junction \cite{gold,gold2}, in the electrochemical oscillators in higher voltage\cite{kiss,kiss2,ott}, in neuronal networks with learning and network adaption\cite{seli}-\cite{niyo}. CS is the most outstanding feature of this  higher harmonic coupling Kuramoto model(HHC-KM).

Here we will investigate HHC-KM from the point of symmetry, and provide a group transformation, and give CS a thoroughly novel interpretation, and answer the question on the same threshold for CS in different parameters $m$.

\section{The generalized Kuramoto model and The transformation to explain CS}

The generalized Kuramoto model with the higher harmonic coupling is
\begin{eqnarray}
\label{Kuramoto mth}
&&\dot{\theta_n}=\omega_n+\frac{K}{N}\sum_{j=1}^{N}\sin m(\theta_j-\theta_n) .
\end{eqnarray}
In the case of small strength $K<K_c$, the term $\omega_n$ dominates the change of the phase $\theta_n$ and the whole phase system is in the incoherent  state. Whenever $K$ exceeds $K_c$, the second terms in Eq.(\ref{Kuramoto mth}) predominate and CS emerges\cite{ott}-\cite{koma2}. It has been also known that CS is sensitive to the initial conditions of the oscillators in Ref.\cite{ott}.

We study CS from completely new view. We will try to find the relation between the generalized and standard Kuramoto models, and penetrate  the phenomena of CS  to study their essence. In the standard Kuramoto model
\begin{eqnarray}
\label{Kuramoto ori}
&&\dot{\theta_n}=\omega_n+\frac{K}{N}\sum_{j=1}^{N}\sin (\theta_j-\theta_n) ,
\end{eqnarray}
the coupling strength $K>0$ is assumed.
By introduction of the transformation
\begin{eqnarray}
\label{trans}
\phi=m\theta,
\end{eqnarray}
together with $m\omega_n,\ mK$ chang into $\omega_n,\ K$,
Eq.(\ref{Kuramoto mth}) takes the form
\begin{eqnarray}
\label{Kuramoto 1-nth}
&&\dot{\phi_n}=\omega_n+\frac{K}{N}\sum_{j=1}^{N}\sin (\phi_j-\phi_n) ,
\end{eqnarray}
 which is the same as that of the standard Kuramoto model\footnote{After completing our work, we notice Ref.\cite{niyo} has a similar transformation for $m=2$ case in the fast study model.}.

 The transformation (\ref{trans}) is crucial to obtain the information on Eq.(\ref{Kuramoto mth}) and give the explanation to CS.  For Eq.(\ref{Kuramoto 1-nth}), the density function $f(\phi,\omega,t)$ in the large $N$ limit satisfies  the continuous equations
\begin{eqnarray}
&&\partial_tf+\partial_{\phi}\bigg[\bigg(\omega+\frac{K}{2i}\bigg(Re^{-i\phi}-  R^*e^{i\phi} \bigg)\bigg)f\bigg]=0,\label{f for trans}\\
&& R=\int \int  f(\phi,\omega,t)e^{i\phi}d\phi d\omega.\label{f for order}
\end{eqnarray}
Generally, the dynamical information for CS is obtained through solve Eqs.(\ref{f for trans})-(\ref{f for order}). Nevertheless, the transformation (\ref{trans}) make it possible to alternatively investigate CS without assorting to the direct solutions to Eqs.(\ref{f for trans})-(\ref{f for order}). See details in the following.

\begin{center}
\begin{figure}[ht]
\begin{tabular}{cc}
\includegraphics[height=0.20\textheight, width=0.45\textwidth]{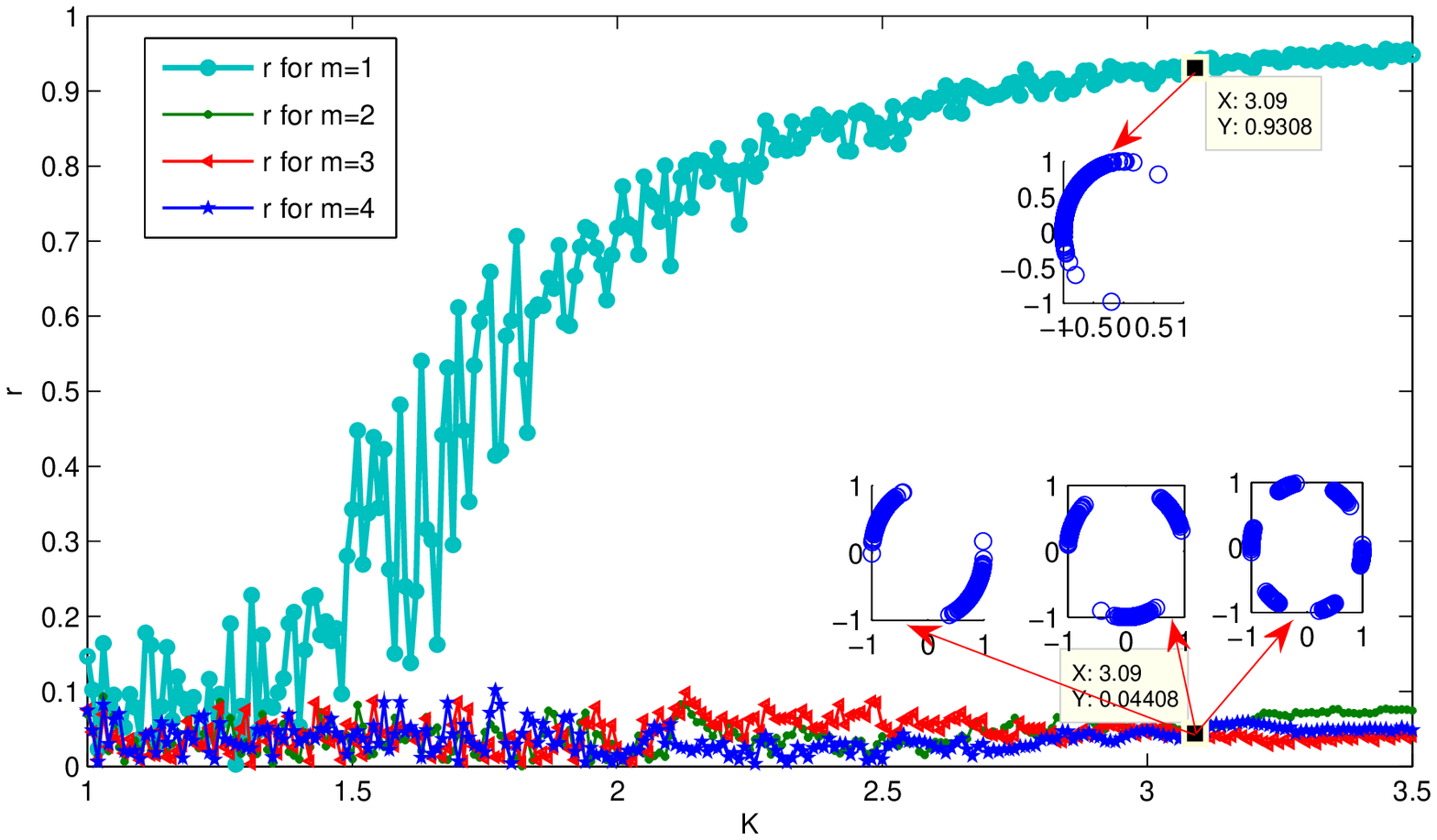}\\
\includegraphics[height=0.20\textheight, width=0.45\textwidth]{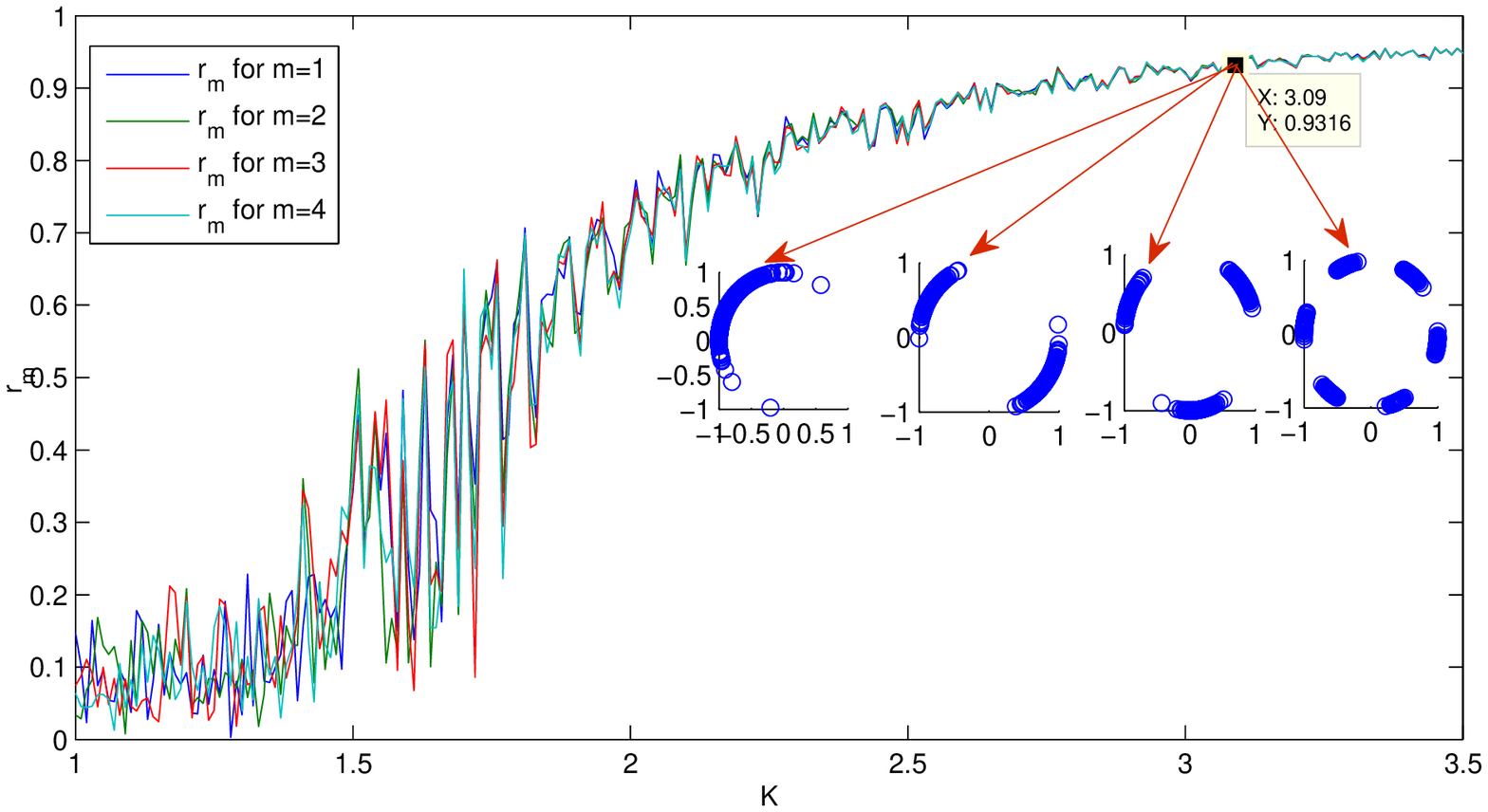}
\end{tabular}
\caption{The order parameter $r(K)$  and $r_m(K)$ against the coupling $K$ for different  $m=1,2,3,6$ with all initial phases distributing uniformly in $(0,2\pi)$.  In the above, the cluster synchronization  appears as $K>K_c\approx 1.55 $ and is shown by the attached three panels for $K=3.09$. Note the order parameters $r\approx 0 $ for $m=2,3,6$ in $K=3.09$, where CS appears, which means $r$ only works well in the case $m=1$.
In the below,  $r_m(K)$  are almost the same for different $m=1,2,3,6$  and indicate the same threshold $K_c$ for the transition from incoherent state to partially synchronized ones. CS is shown for $K=3.09$ with the oscillators'positions  indicated by the small blue circles in the corresponding large circles  for $m=1,2,3,6$ respectively. The four large circles (circle's lines are not shown) are attached into the figure. There are $500$ oscillators and their positions are indicated by the small blue circles in the corresponding circles for $m=1,2,3,6$
respectively.}\label{fig1}
\end{figure}
\end{center}

 Suppose initially uniform distribution in $(0,2\pi)$ for the phases $\theta_n, n=1,2,\cdots, N$, the transformation indicates the corresponding initial phases' uniform distribution is in $(0,2m\pi)$ for the phases $\phi_n, n=1,2,\cdots, N$. Because Eq.(\ref{f for trans}) is periodic in $\phi$, with the initial periodic condition in $\phi$, the solution $f(\phi,\omega,t)$ is also periodic in $\phi$.  So one has \[f(\phi,\omega,t)=f(\phi+2\pi,\omega,t)=\cdots=f(\phi+2(m-1)\pi,\omega,t),\]
 which results in  the following outstanding properties for the corresponding density function $f(\theta,\omega,t)$
 \begin{eqnarray}
 f(\theta,\omega,t)&=& f(\theta+\frac{2\pi} m,\omega,t)\nonumber\\
 &=& \vdots \nonumber\\
 &=& f(\theta+\frac{2(m-1)\pi}{m},\omega,t).\label{symmetry in theta}
 \end{eqnarray}
 Hence the cluster phenomenons appear, and the phases  $\theta_n, n=1,2,\cdots, N$ cluster into $m$ sections. From Eq.(\ref{symmetry in theta}), it is easy to see that
 the order parameter is zero no matter  the phases $\theta_n, n=1,2,\cdots, N$ are in CS state or not, that is,
\begin{eqnarray}
\label{order in theta}
&&re^{i\Psi}=\int_{-\infty}^{\infty} \int_0^{2\pi}  f(\theta,\omega,t)e^{i\theta}d\theta d\omega=0.
\end{eqnarray}
So, the order parameter no longer works in the generalized one, as is shown in Fig.\ref{fig1} for the cases of $m=2,\ 3,\ 6$ and is substituted by the generalized  order parameter $r_m$ defined as
 \begin{eqnarray}
\label{order m in theta}
&&R=r_me^{i\Psi'}=\frac{1}{N}\sum_{j=1}^{N} e^{im\theta_j}, or\nonumber\\
&&R=r_me^{i\Psi'}=\int_{-\infty}^{\infty} \int_0^{2\pi}  f(\theta,\omega,t)e^{im\theta}d\theta d\omega
\end{eqnarray}
 Eq.(\ref{symmetry in theta})   guarantees  the generalized order parameters  $r_m$ being
 the same for all parameters $m$, which also could be obtained from the fact of the same density function $f(\phi,\omega,t)$ for different $m$ in calculation $r_m=|R|$ by Eq.(\ref{f for order}).
 We also numerically calculate $r_m$ against $K$ forwardly for different $m=1,2,3,6$ with the same initial random distributions in $(0,2\pi)$ and   the numerical results confirm the conclusion. See the second panel in Fig.\ref{fig1} for detail.

\begin{center}
\begin{figure}
\begin{tabular}{cc}
\includegraphics[height=0.15\textheight, width=0.21\textwidth]{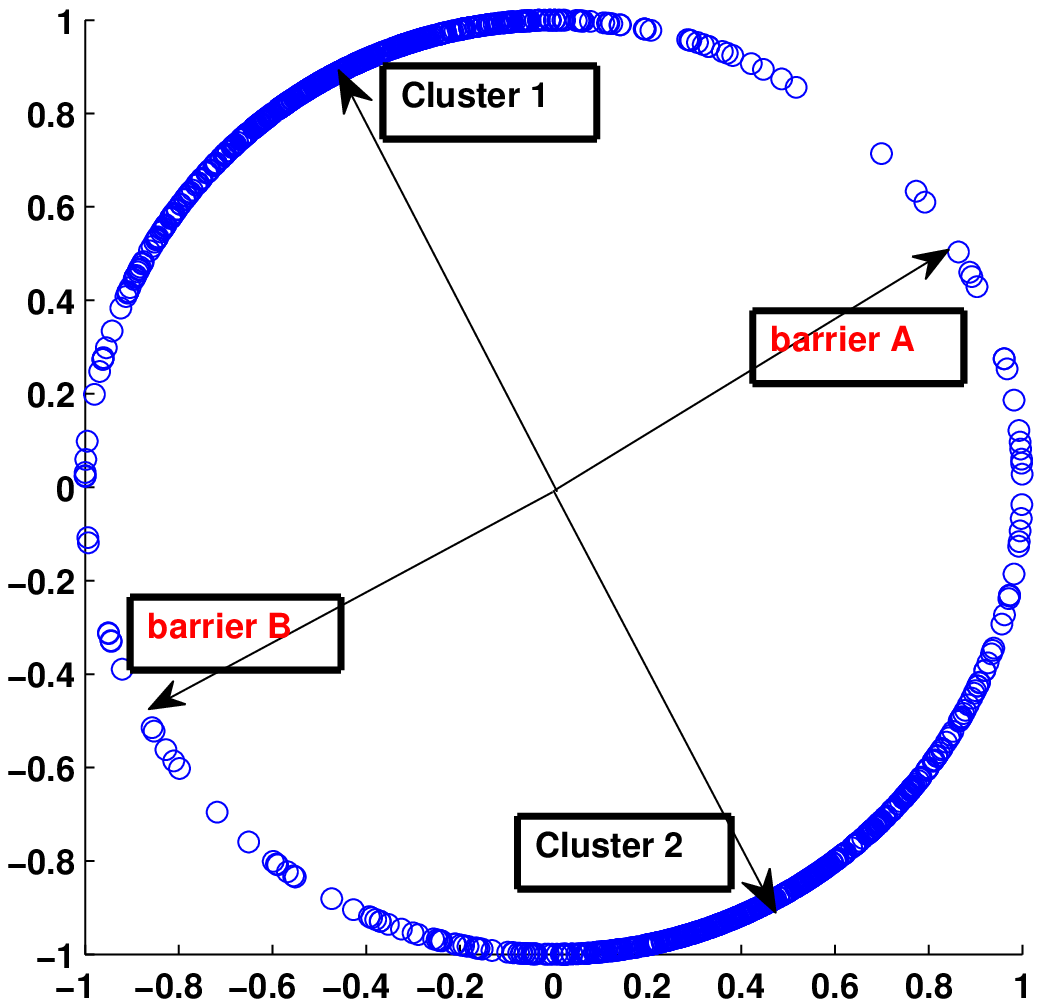}
\includegraphics[height=0.15\textheight, width=0.21\textwidth]{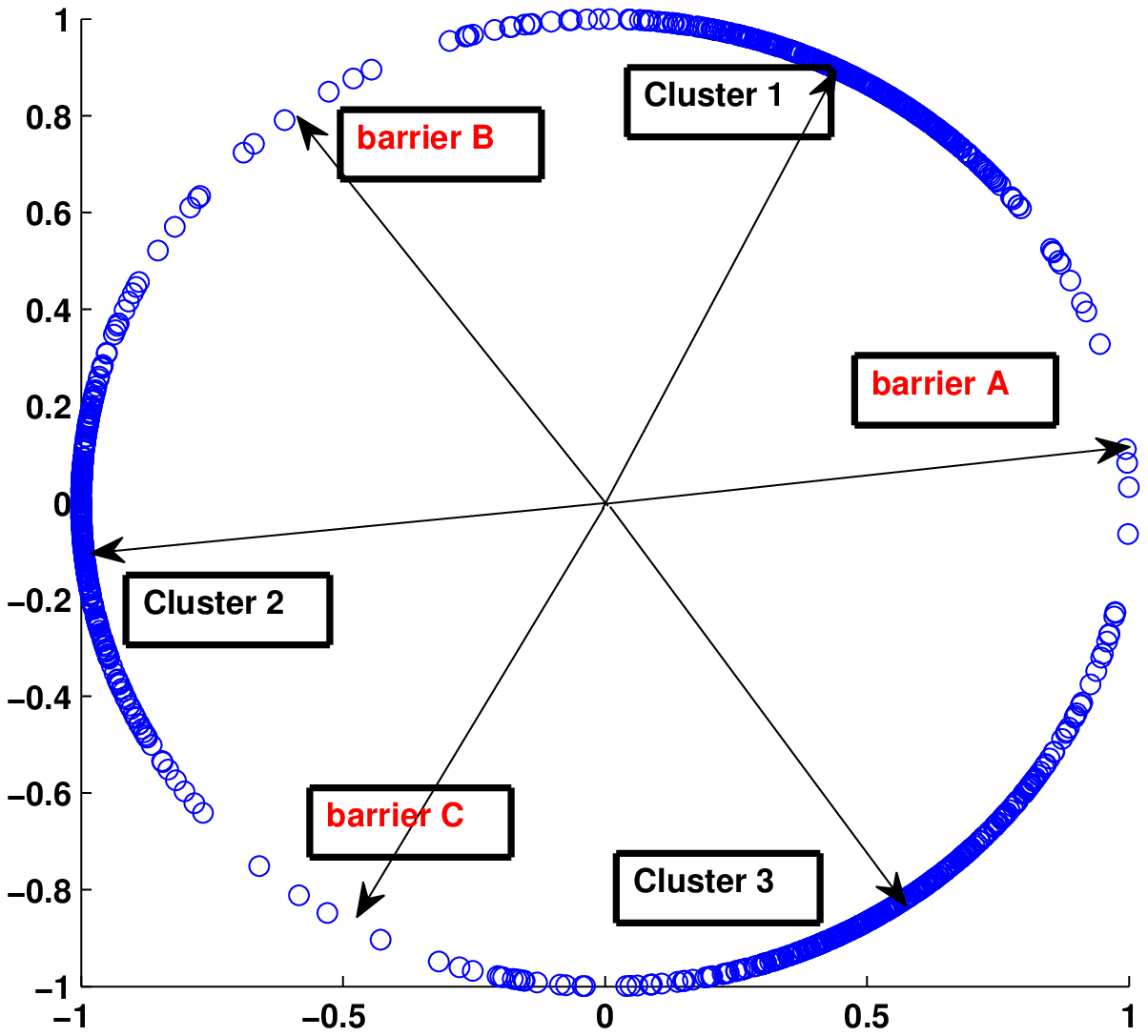}
\end{tabular}
\caption{ synchrony clusters have their centers at $\alpha-\frac{n\pi}{m}$  and  their boundaries $\alpha+\frac{(2n-1)\pi}{2m}, \alpha+\frac{(2n+1)\pi}{2m}$ for $ \ n=0,\ 1, \cdots,\  m-1$, which are shown by red arrows in the big circles. The final evolutionary positions of the oscillators  on the circles are denoted by blue small circle.  The synchrony oscillators could not pass through the boundaries, which act like potential barriers, while the asynchrony oscillators have enough energies (higher or lower natural frequencies) to overcome the barriers and are not confined in one section. The  parameters are $K=2,\ m=2,3$ in  the two panels.}
\label{fig2}
\end{figure}
\end{center}

 Note another symmetry of Eq.(\ref{Kuramoto mth}), that is, under the translation
 \begin{eqnarray}
 \theta_n\rightarrow \theta_n+\bar{\alpha},\ \forall 1\le n \le N,\label{translation}
 \end{eqnarray}
 Eq.(\ref{Kuramoto mth}) is unchanged.
 So the cluster sections might be $(\alpha-\frac{\pi}{2m},\alpha+\frac{\pi}{2m}),\  (\alpha+\frac{\pi}{2m},\alpha+\frac{3\pi}{2m}),\  \cdots$, as are shown in Fig.\ref{fig1} and Fig.\ref{fig2}.
 These cluster sections naturally have boundaries, which separate the different cluster sections.  The boundaries of the cluster section with its center at $\alpha+\frac{n}{m}\pi, n=0, 1,2,\cdots, m$ are $\alpha+\frac{(2n-1)\pi}{2m},\ \alpha+\frac{(2n+1)\pi}{2m}$. The most important feature of the boundaries is their potential-barrier characteristic: after the formation of CS, the synchrony phases in each section can only stay in its section, only the asynchrony phases do pass the barriers. The remarkable properties also come from the normal Kuramoto system combining with the transformation Eq.(\ref{trans}). See details in the following.

 In the normal Kuramoto system (\ref{Kuramoto 1-nth}), the synchrony state forms around its center $\beta$ (we define its center' angle as $\beta$), and the synchronization oscillators will stay in the section $(\beta-\frac{\pi}2, \beta+\frac{\pi}2)$. Hence the boundaries of synchrony state lie at $\beta-\frac{\pi}2$ and $\beta+\frac{\pi}2$. The oscillators whoever already are synchronized can not go cross the boundaries, so the boundaries behave as  potential barriers to forbid the synchronized oscillators to pass through.
Nevertheless, the oscillators not synchronized will have enough 'energy' (high positive or negative frequency) to overcome the barriers and go beyond them.
From the periodic properties of the oscillators' phases, the above synchrony state could be regarded as the sections in $(\beta+2n\pi-\frac{\pi}2, \beta+2n\pi+\frac{\pi}2),\ n\in \mathfrak{N}$, which actually are the same section for phases $\phi_j, j=1,2,\cdots, N$. $m$ ones of these sections
with $n=0,1,,2,\cdots, m-1$ are the same from the phases of $\phi_j, j=1,2,\cdots, N$, nevertheless, they will be completely different sections when they are transformed back to the phases $\theta_j, j=1,2,\cdots, N$ with each one denoted by $(\alpha+\frac{(2n-1)\pi}{2m}, \alpha+\frac{(2n+1)\pi}{2m})$ for $n=0,1,,2,\cdots, m-1$. In the same way, the boundaries of each section also are potential barriers for phases, as already stated in above, see Fig.\ref{fig2}. In extreme case, the  coupling strength $K$ is so large that no phase will have enough energy to overcome the boundary barriers, as in the numerical simulation the maximum of frequency is limited. So all phases will synchronized into one of the  $m$ sections and no one can get over the boundary barriers, as shown in Fig.\ref{fig1}. In the following, we will confine our discussion in this special case to discuss that CS is sensitive to the initial conditions, which is numerical shown in Fig.\ref{fig4}.

\begin{center}
\begin{figure}
\begin{tabular}{ccc}
\includegraphics[height=0.15\textheight, width=0.21\textwidth]{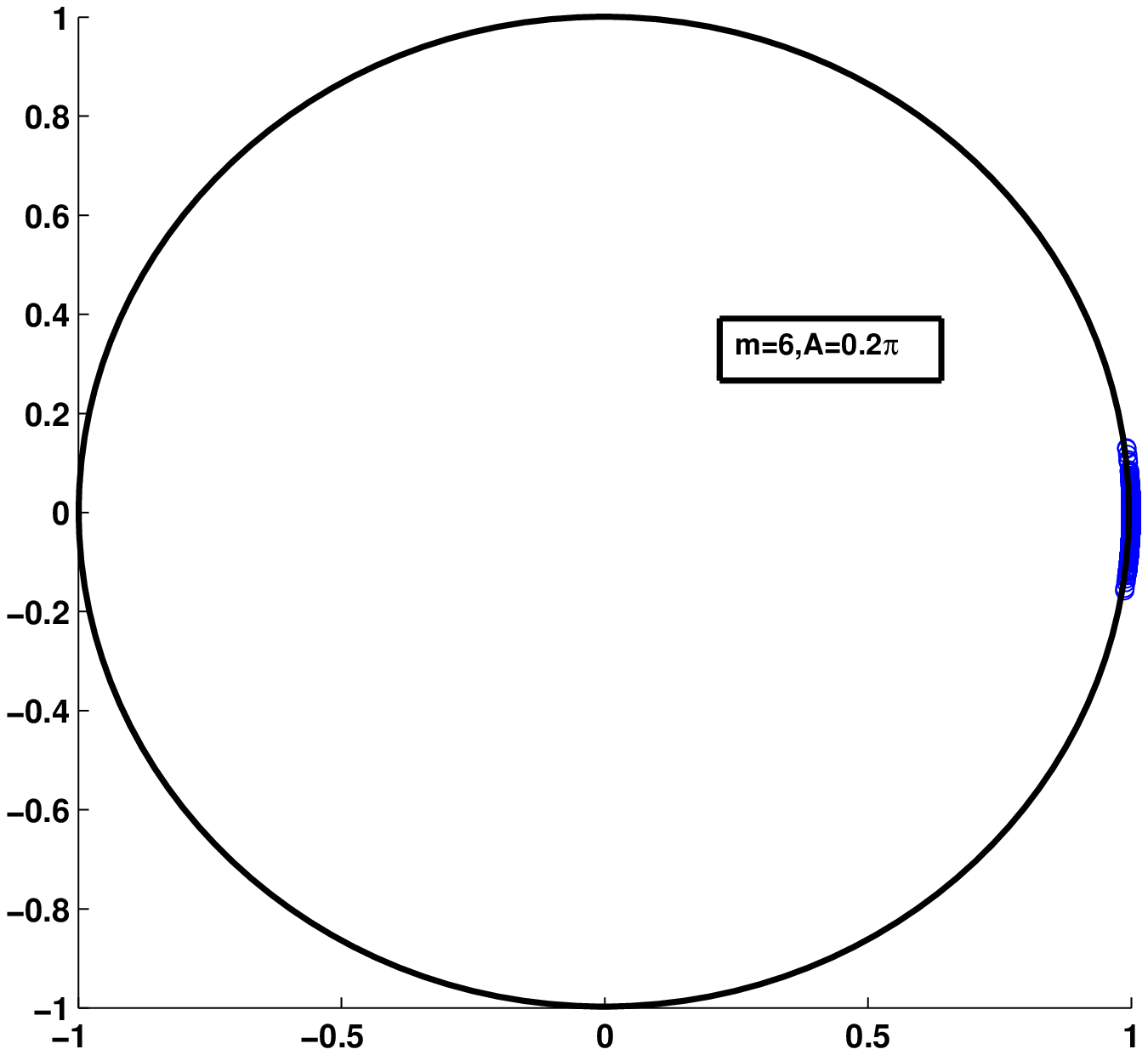}
\includegraphics[height=0.15\textheight, width=0.21\textwidth]{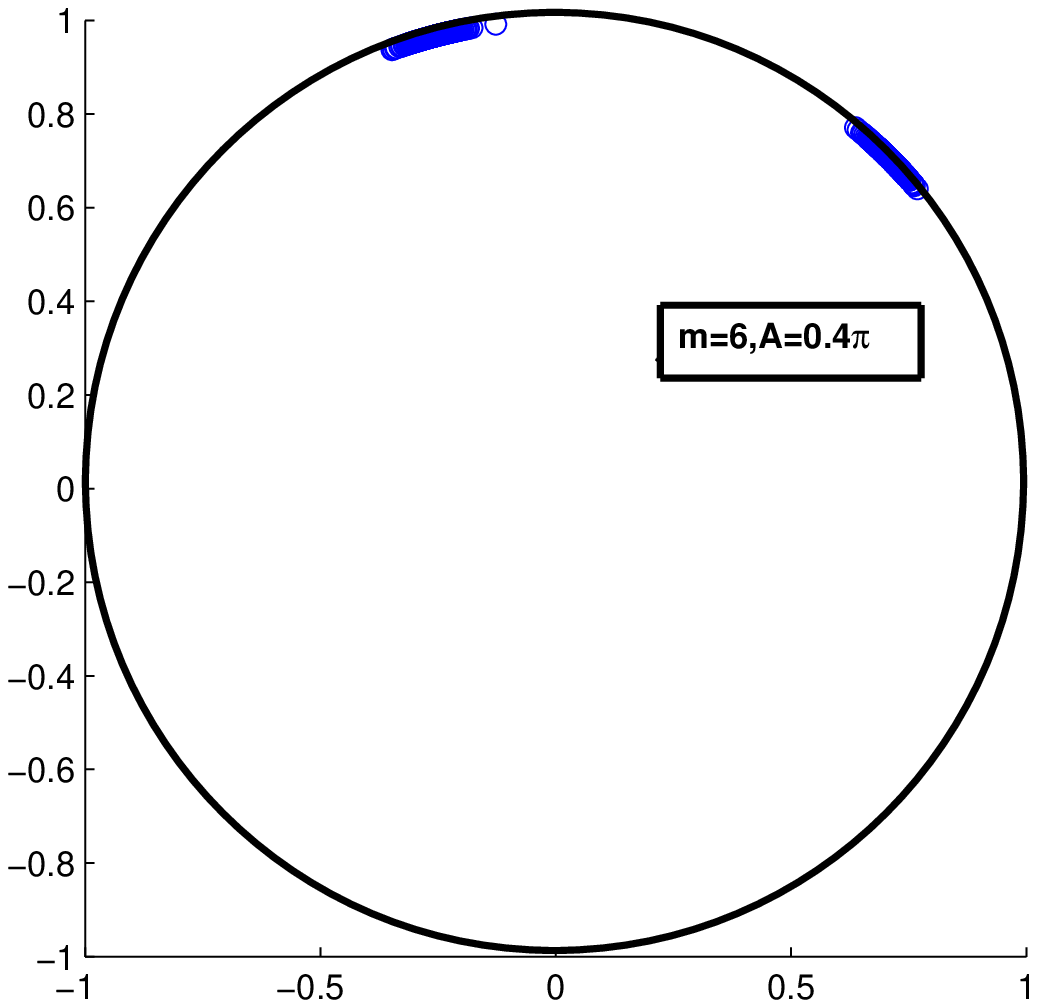}
\end{tabular}
\caption{Schematic diagrams on the sensitiveness of CS to $A$ for $A=0.2\pi, 0.4\pi$. (the initial uniform phases distributions are in $(0,A)$ ). The same parameters $K=5,\ m=6$ in all two panels. From $0.2\pi<\frac{n\pi}{6}$ for $n=1$ and $\frac{(n-1)\pi}{6}<0.4\pi<\frac{n\pi}{6}$ for $n=2$, it is easy to see that one CS section forms in the left panel and two CS sections form on the right panel.}
\label{fig3}
\end{figure}
\end{center}

\begin{center}
\begin{figure}
\begin{tabular}{cc}
\includegraphics[height=0.15\textheight, width=0.21\textwidth]{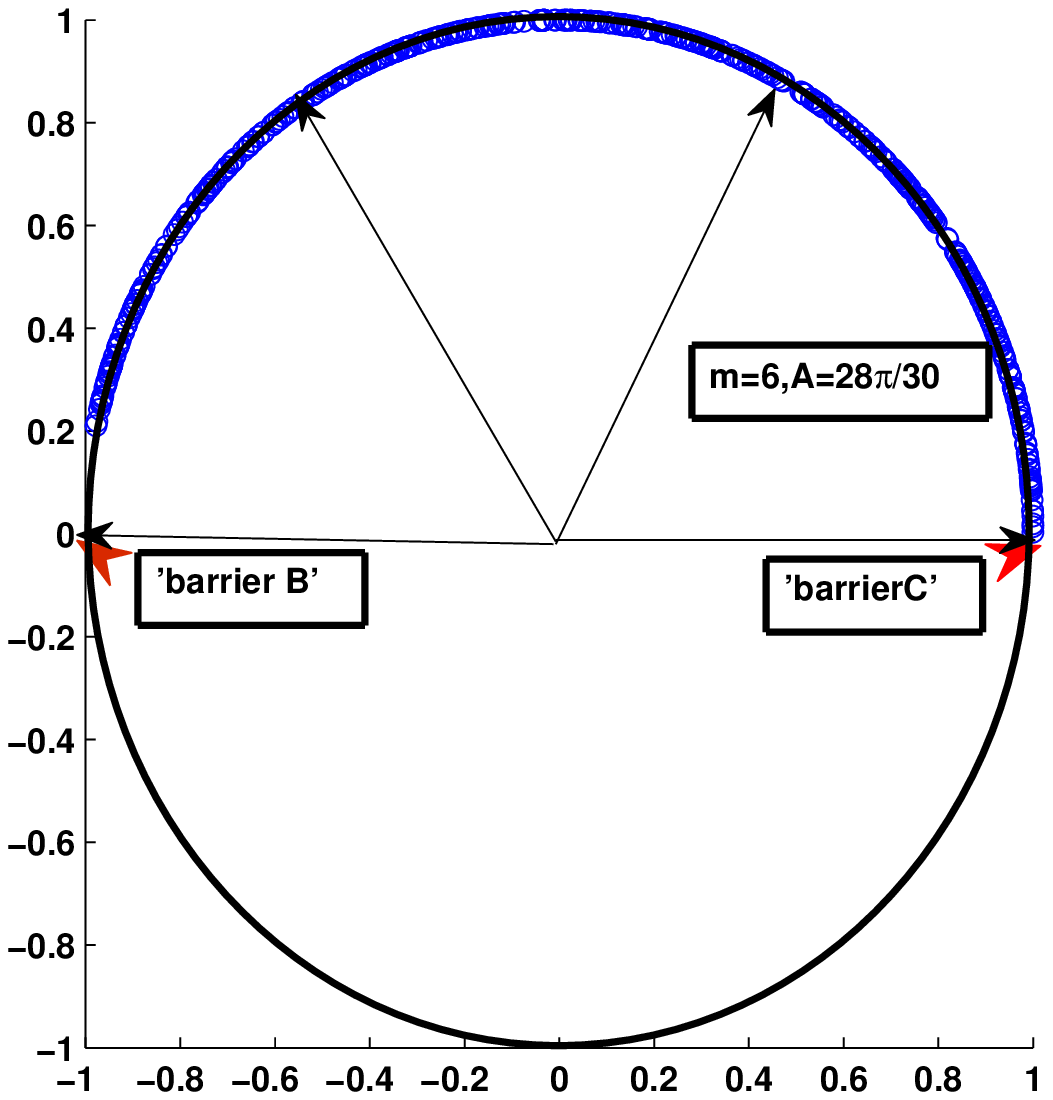}
\includegraphics[height=0.15\textheight, width=0.21\textwidth]{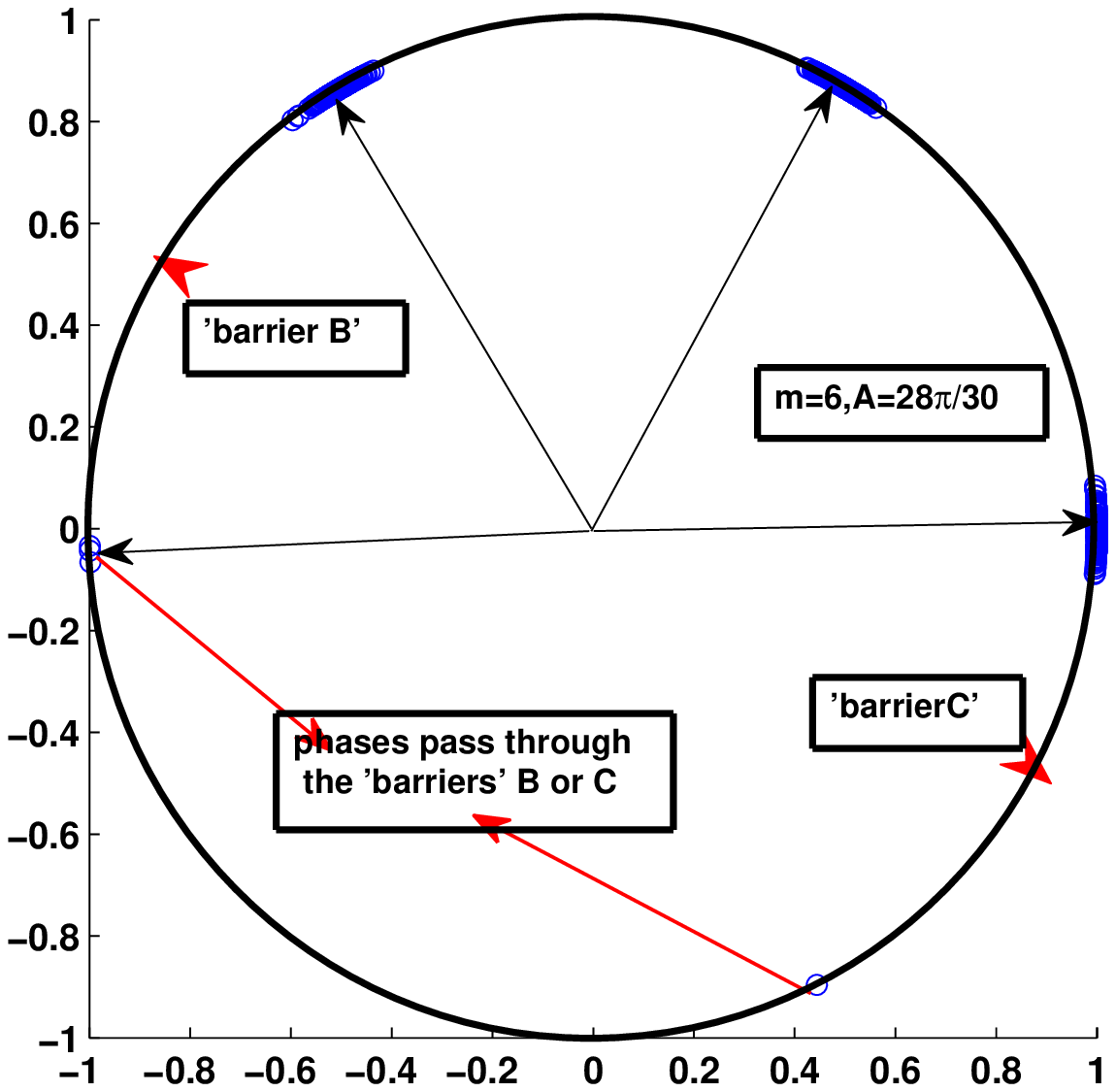}\\
\includegraphics[height=0.15\textheight, width=0.21\textwidth]{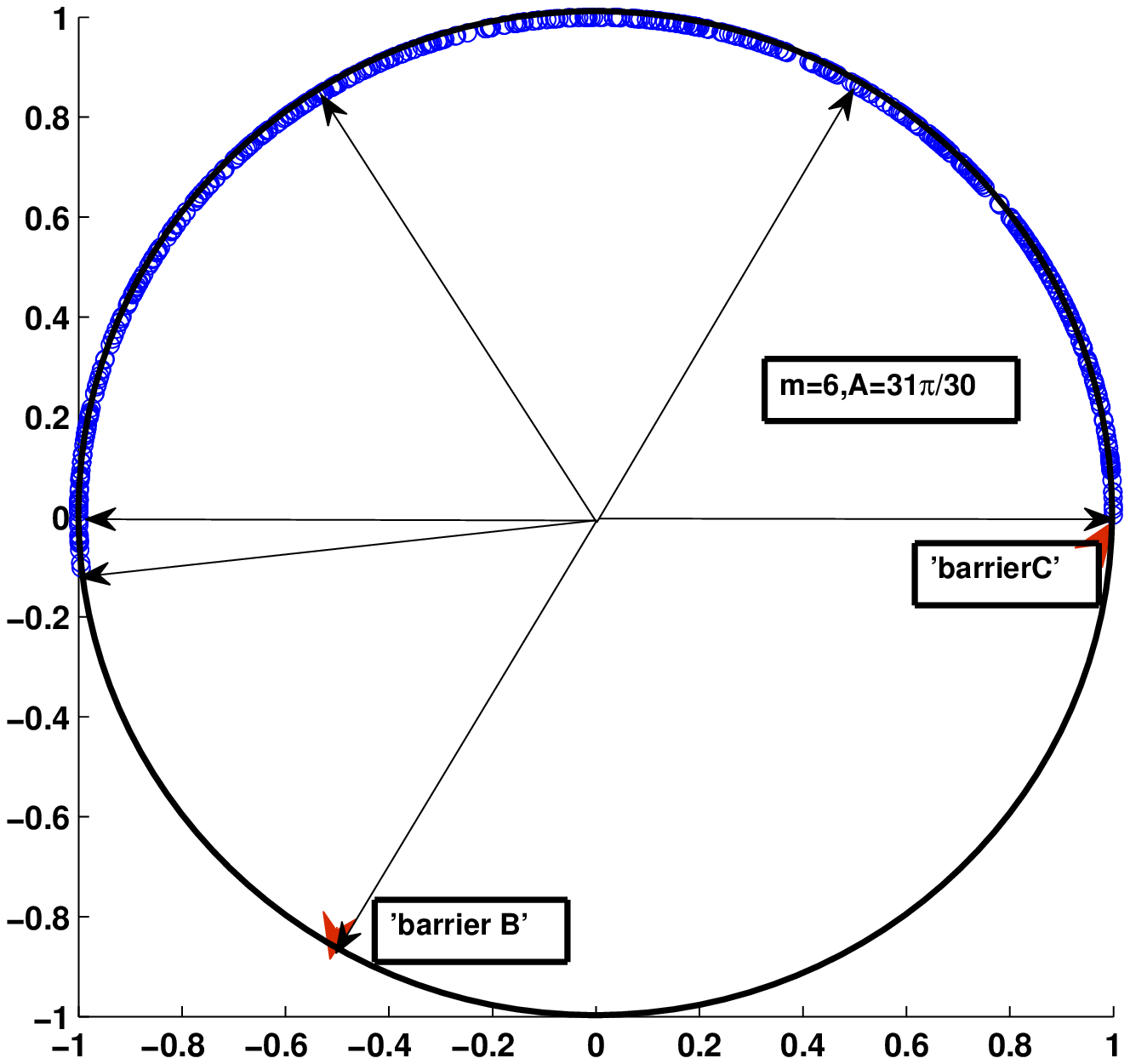}
\includegraphics[height=0.15\textheight, width=0.21\textwidth]{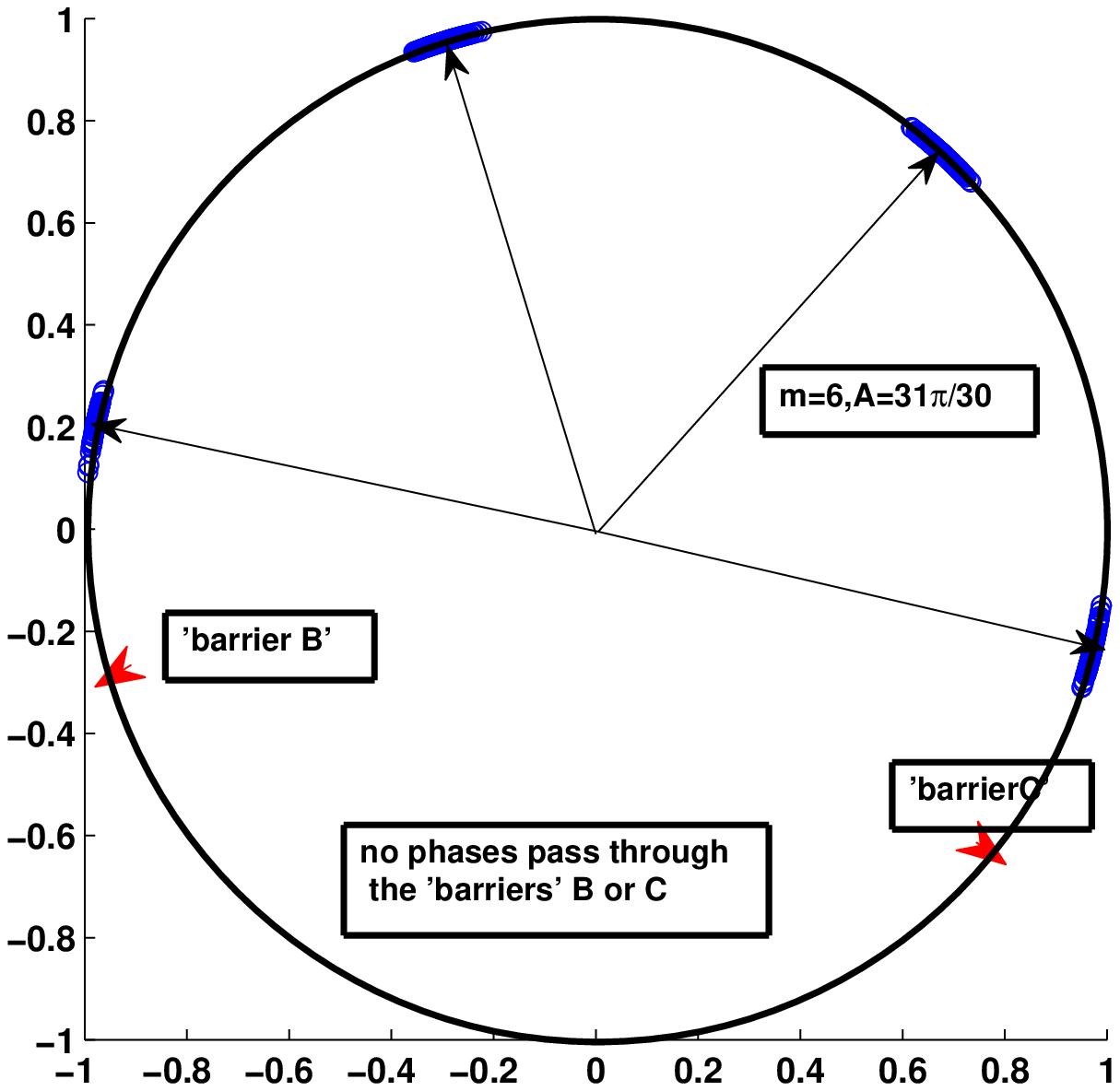}
\end{tabular}
\caption{Initial random distribution in $(0,A) for A=\frac{28\pi}{30},\ A=\frac{31\pi}{30}$ in the left panels, and their final distribution are shown by the tiny blue circles on the big circle in the right panels.  The  parameters are $K=5,\ m=6$ in all the panels. Only several oscillators (actually $4$ oscillators) pass through the boundaries' barriers $B$ or $C$, denoted in the upper right panel. There are altogether 500 oscillators in each circle.}
\label{fig4}
\end{figure}
\end{center}

The periodic property for $f(\phi,\omega,t)$ might be violated by the initial condition of the phases. in this case, cluster phenomenons also are destroyed somehow and show CS is sensitive to the initial condition, as former investigation indicated\cite{ott}, See Fig.\ref{fig3} for details.
As stated above, we suppose $K\gg K_c$. The initial distribution falls into  $(0, A)$  with  $\frac{2(n-1)\pi}m <A<\frac{2n\pi}m,\ n<m$. It can be supposed  that there are initially about $n$ sections, so the boundaries will prevent all the oscillators except ones on the boundaries to pass through. Hence, the oscillators will evolve into $n$ cluster sections plus very small part of the oscillators enters into the $(n+1)-th$ section, see Fig.\ref{fig4}.
However, Whether the number of cluster sections is $n$ or $(n+1)$ is very sensitive to initial conditions of the oscillators. For example, it is possible to form $(n+1)$ cluster sections if there are many oscillators near the boundaries $B$ or $C$, as is the case shown on the second and the fourth panels in Fig.\ref{fig4}.

\section{Conclusion and discussion}

CS  has been investigated by the method of self-consistent approach in Refs.\cite{ott, koma}, \cite{seli}-\cite{ golo}. Neural network actually studied the combination of the first and second harmonic couplings in the generalized Kuramoto model\cite{seli}-\cite{golo}, which is also treated  in Ref.\cite{koma}, \cite{koma2}.
In the $N$ identical oscillators'case, the symmetry viewpoint is applied and CS of the two groups of $m$ and $N-m$ oscillators   is connected with their symmetry groups of the dynamics $S_m\times S_{N-m}$ \cite{Ashwin}, \cite{ban}. The symmetry group $S_N$ is only suited for the identical oscillators in the Kuramoto model. However, it is still very difficult to obtain clear analytical results by the self-consistent approach and detailed understanding of CS \cite{koma, koma2}.

In the nice work \cite{ott}, CS have been investigated by the self-consistent approach.  The density function for the second harmonic coupling case is decomposed into the symmetric and asymmetric parts in Ref.\cite{ott} , and the Ott-Antonsen (OA) mechanism is utilized to analyze the symmetric case. However, the asymmetric one is not accessible to the analytical study, and numerical methods are needed to the full solution of the density function \cite{ott}. For higher harmonic coupling than the second, the density function $f(\theta,\omega, t)$ is decomposed in to $m$
parts as $f=f^{(1)}+\cdots +f^{(m)}$ and $f^{(j)}=\sum_{n=-\infty}^{\infty}a^{(j)}_ne^{i(m*n+j)\theta},\ j=1,2,\cdots, m$. OA mechanism could be utilized for $f^{(m)}$ and the critical strength $K_c=2\Delta$ is obtained for the the Lorentz's distribution of the natural frequency $g(\omega)=\frac{\Delta}{\pi (\omega^2+\Delta^2)}$. However, it is not easy to obtain other $f^{(j)},\ j\ne m$ and numerical methods are used  for $f(\theta,\omega,t)$\cite{ott}.

However, our study is completely different  from that in Ref.\cite{ott}. We mainly rely on the transformation (\ref{trans}) and the periodic properties of the density function to study the most typical phenomena CS in the generalized Kuramoto model.  By the transformation (\ref{trans}), it is possible to relate CS with the periodic properties of the density function $f(\phi,\omega, t)$ or $f(\theta,\omega, t)$. To hold the periodic properties for $f(\theta,\omega, t)$, the initial distribution of the oscillators in terms of $\theta_n, n=1,2, \cdots, N$ must range randomly in $(0,2\pi)$. Because of the periodic properties of $f(\phi,\omega, t)$ or $f(\theta,\omega, t)$ and the relation between $\theta$ and $\phi$, the $m$ cluster synchrony  states appear corresponding to $\theta$, which are in one of the sections $(\alpha+\frac{(2n-1)\pi}{2m}, \alpha+\frac{(2n+1)\pi}{2m})$ for $n=0,1,,2,\cdots, m-1$.  Corresponding the cluster sections, there naturally exist boundaries for them, which function as the potential barriers to forbid the synchrony oscillators to pass through. The existence of the barrier-like boundary can also explain the sensitiveness of CS to the initial conditions.
The initial distribution of the phases $\theta$ in $(0,A)$ with $A<2\pi$ will break the periodic condition for $f(\phi,\omega, t)$ or $f(\theta,\omega, t)$ and the violation  will result in the sensitiveness of CS to the initial distribution (0,A).
The explanation to CS in the letter is novel and simple, and has both the profound mathematical insight and  clear physical understanding. Our detailed numerical studies confirm the symmetric analysis.
\acknowledgments
The work was partly supported by the National Natural Science of China (No. 10875018)
and the Major State Basic Research Development Program of China (973 Program: No.2010CB923202).


\begin{thebibliography}{99}
\bibitem{kura}
 Kuramoto Y., Chemical Oscillations, Waves, and Turbulence (Springer, Berlin,1984).
\bibitem{aceb}Acebron J. A. , Bonilla L. L. , Perez Vicente C. J. ,  Ritort F., and Spigler R., Rev. Mod. Phys. {77} {2005} {137}.
\bibitem{stog}Strogatz S. H.,
Physica D  {143} {2000} { 1}.
\bibitem{kura ni} Kuramoto Y. and  Nishikawa I., J. Stat. Phys. {49} {1987} { 569}.
\bibitem{gupta}Gupta S., Campa A., and Ruffo S.,
J. Stat. Mech.: Theory Exp. {R08001} {2014}  { 1}.
\bibitem{tana}Tanaka H., Lichtenberg A. J. and Oishi S.,
Phys. Rev. Lett. {78}  {1997} ({2104}.
\bibitem{aceb2}Acebron J. A. and Spigler R.,
Phys. Rev. Lett. {81} {1998}{2229}.
\bibitem{aceb3} Acebron J. A., Bonilla L. L. and Spigler R.,  Phys. Rev. E {62} {2000} {3437}.
\bibitem{wang}  Wang H. and Li X., Phys. Rev.E. {83} {2011}{ 066214}.
\bibitem{zhang} Zhang X., Hu X., Kurths J., and Zonghua Liu., Phys. Rev.E. {88} {2013} { 010802(R)}.
\bibitem{rafa}   Pinto Rafael S. and  Saa A., Phys.Rev. E {91} {2015}{022818}.
\bibitem{zou}  Zou Y., Pereira T. ,  Small M., Liu Z., and  Kurths J., Phys. Rev. Lett. {112} {2014} { 114102}.
\bibitem{juses}  Gomez-Gardenes J.,  Gomez S.,  Arenas A., and  Moreno Y., Phys. Rev. Lett. {106} {2011} { 128701}.
\bibitem{abram} Abrams D. M. andStrogatz  S. H.,  Phys. Rev. Lett.,{93} {2004} {174102}.
\bibitem{fu}  Fu C.,  Deng Z.,  Huang L., and Wang X.,  Phys. Rev. E {87} {2013} { 032909 }.
\bibitem{wu}  Wu Y., Xiao J., G. Hu  and M. Zhan, Europhys. Lett. {97} {2012} { 40005}.
\bibitem{yang} Zhu Y., Zheng Z. and Yang J., Phys. RevE.{89} {2014} { 022914}.
\bibitem{ping} Ju P., Dai Q., Cheng H., and Yang J., Phys.Rev.E {90} {2014} {  019903}.
\bibitem{ott}   Skardal P. S.,  Ott E. and  Restrepo J. G., Phys. Rev. E {84} {2011} { 036208}.
\bibitem{koma}   Komarov M. and  Pikovsky A., Phys. Rev. Lett. PRL {111} {2013} {204101}.
\bibitem{gold}  Goldobin E.,  Koelle D.,  Kleiner R. and R. G. Mints.,  Phys. Rev. Lett. {107} {2011} { 227001} .
\bibitem{gold2}  Goldobin E.,  Kleiner R., Koelle D. and R.G. Mints., Phys. Rev. Lett. {111} {2013} {057004} .
\bibitem{kiss}   Kiss I. Z., Zhai Y. and  Hudson J. L., Phys. Rev. Lett. {94} {2005} {248301}.
\bibitem{kiss2}  Kiss I. Z., Zhai Y. and  Hudson J. L.,Prog. Theor. Phys. Suppl. {161} {2006} { 99}.
\bibitem{seli}  Seliger P.,  Young S. C. and  Tsimring L. S., Phys. Rev. E {65} {2002} { 041906}.
\bibitem{Hansel1}Hansel D.,  Mato G. and  Meunier C.,  Europhys. Lett. {23} {1993} {367}.
\bibitem{Hansel3}Hansel D.,  Mato G. and  Meunier C.,  Phys. Rev. E {48} {1993}
{3470}.
\bibitem{Ashwin}Ashwin P. and  Borresen J., Phys. Rev. E {70} {2004}{026203}.
\bibitem{ban}Banaji M.,  Phys. Rev. E {71} {2005} { 016212}.
\bibitem{niyo}Niyogi R. K. and  English L. Q., Phys. Rev. E {80} {2009} { 066213}.
\bibitem{okud}Okuda K., Physica D {63} {1993}  {424}.
\bibitem{golo}Golomb D.,  Hansel D.,  Shraiman B. and  SompolinskyH., Phys. Rev. A {45} {1992} {3516}.
\bibitem{koma2}Komarov M. and  Pikovsky A., arXiv:1404.7292 {2014}.
\bibitem{otta}Ott E. and  Antonsen T. M., Chaos {18} {2008} { 037113}.
\bibitem{Pikovsky1} Rosenblum M. and  Pikovsky A., Phys. Rev. Lett. {98} {2007} {064101}.
\bibitem{Pikovsky2} Pikovsky A. and  Rosenblum M., Physica D: Nonlinear Phenomena {238} {2009} { 27} .
\bibitem{Pikovsky5} Baibolatov Y.,  Rosenblum M.,  Zhanabaev Z. Z.,  Kyzgarina M. and  Pikovsky A., Phys. Rev. E {80} {2009} {046211}.
\bibitem{tian}G. Tian, S. Hu, S. Zhong, Arxiv:1505.03660, (2015)
\end{thebibliography}
\end{document}